\documentstyle[amssymb,12pt]{article}

\input{tcilatex}
\begin{document}

\begin{titlepage}
\begin{center}
{\hbox to\hsize{
\hfill \bf HIP-2002-33/TH }}
{\hbox to\hsize{\hfill July 2002 }}

\bigskip
\vspace{3\baselineskip}

{\Large \bf

Kaluza-Klein decomposition and gauge coupling unification in orbifold GUTs\\}

\bigskip

\bigskip

{\bf Masud Chaichian$^{\mathrm{a}}$ and Archil Kobakhidze$^{\mathrm{a,b}}$ \\}
\smallskip

{ \small \it
$^{\mathrm{a}}$High Energy Physics Division, Department of Physical Sciences, 
University of Helsinki $\&$ \\
Helsinki Institute of Physics, FIN-00014 Helsinki, Finland\\ 
$^{\mathrm{b}}$Andronikashvili Institute of Physics, Georgian Academy of Sciences, 
GE-380077 Tbilisi, Georgia\\}

\bigskip

\vspace*{.5cm}

{\bf Abstract}\\
\end{center}
\noindent
We discuss Kaluza-Klein (KK) decomposition in 5-dimensional (5D) field theories 
with orbifold compactification. Kinetic terms localized at orbifold fixed points, 
which are inevitably present in any realistic model, modify the standard KK mass spectrum and   
interactions of KK modes. This, in turn, can significantly affect phenomenology 
of the orbifold models. As an example, we discuss gauge coupling unification in N=1 supersymmetric 
5D orbifold SU(5) model. We have found that uncertainties in the predictions of the model 
 related to modification of the KK masses are large and essentially uncontrollable.

\bigskip

\bigskip

\end{titlepage}

\section{Introduction}

Almost all currently available experimental data in particle physics are
described by the Standard Model with an impressive accuracy. However, it is
widely believed that various theoretical puzzles and problems of the
Standard Model indicate that the theory in the present form is just the low
energy limit of a more fundamental theory. Since their early days \cite{1}
Grand Unified Theories (GUTs) are viewed as one of the most attractive
candidates for such a theory. Unification of three gauge couplings within
the Minimal Supersymmetric Standard Model \cite{2} strongly supports the
basic GUT idea and low-energy broken supersymmetry.

Construction of realistic GUTs in 4 dimensions (4D), however, faces certain
difficulties in understanding some phenomenologically important issues such
as proper breaking of GUT gauge symmetry, doublet-triplet hierarchy, proton
stability, correct relation between the masses of quarks and leptons, etc.
Recently some of the problems of conventional 4D GUTs has been addressed in
the context of higher-dimensional field theories with orbifold
compactification [3-13]. Particularly, GUT symmetry breaking can be realized
through the orbifold compactification which could also naturally provide the
doublet-triplet splitting and avoid coloured Higgsino (Higgs) mediated
proton decay. Moreover, by placing matter fields in the higher-dimensional
bulk one automatically obtains the theory with stable (in all orders of
perturbation theory) proton \cite{5} and without SU(5) GUT relations among
the masses of the down quarks and charged leptons which are apparently wrong
for the two lightest families. Alternatively, restricting the third family
of fermions to a SU(5)-invariant orbifold fixed-point, one can keep the
celebrated b-$\tau $ unification \cite{6,7}. Other aspects of fermion masses 
\cite{11} and the possible origin of the observed structure of fermion
families \cite{12,13} have been also widely discussed. Summarizing, it turns
out that higher-dimensional orbifold GUTs could give a simple intrinsically
geometric explanation of various problems of conventional 4D models and
provide a phenomenologically more attractive bottom-up approach than the
highly restricted top-down approach to the unification in higher-dimensions
in the framework of fundamental string theory. Notice, however, that
consistency of higher-dimensional field theories and particularly field
theories on singular spaces, such as orbifolds, requires some ultraviolate
(UV) completion \cite{9}. Such a completion is commonly viewed to be
provided by the fundamental string theory.

An unspecified UV physics generally leads to uncertainties in the effective
low energy field theories. Namely, from the field theory point of view one
can expect appearance of certain operators localized on the orbifold fixed
points \cite{1} which respect symmetries on the orbifold subspaces rather
than the symmetries in the bulk. Among them naturally appear the relevant
''renormalizable'' operators. Particularly, one can write down kinetic terms
of certain orbifold fields on the GUT symmetry breaking fixed point. Since
they do not respect GUT symmetry one can worry that such operators could
completely destroy higher-dimensional gauge coupling unification. The
localized kinetic terms for the gauge fields have been studied in \cite{6,7}
and \cite{8}. It was pointed out in \cite{6,7} that assuming the unification
happens in the strong coupling regime one can dilute the UV sensitivity and
even improve the apparently wrong prediction for the strong gauge coupling
of the 4D supersymmetric SU(5) model. High-precision prediction for the
strong gauge coupling constant obtained in \cite{7} singles out the simple
N=1 supersymmetric 5D SU(5) GUT.

In this paper we would like to reconsider the effects of the kinetic terms
localized on the orbifold fixed points within the 5D effective field
theories. We will show that local kinetic terms modify the KK mode
decomposition and as a result the entire phenomenology of low energy
orbifold field theories will be affected significantly. As an implication to
the orbifold GUT, we have considered the influence of local kinetic terms on
the gauge coupling unification. The essential point is that the localized
kinetic terms for the Higgs and matter fields can not be constrained to be
small by the requirement of strong gauge coupling unification. We have found
that uncertainties related with local kinetic terms of Higgs and matter
fields are generically large and uncontrollable within the effective field
theory approach. In this way the predictions of the model are UV sensitive
due to incalculable parameters related with localized kinetic terms. Within
these uncertainties in the prediction of strong gauge coupling, one can
therefore successfully incorporate different orbifold GUT models as well.

The effects of localized kinetic terms have already been discussed within
different models. Importance of such terms for the quasilocalization of
higher-dimensional gravity and gauge fields in the infinite volume extra
dimensions has been pointed out in \cite{14} (see also \cite{15}). The case
of large compact extra dimension(s) with single brane has been subsequently
studied in \cite{16}. It has been shown in \cite{17} that local kinetic
terms for bulk fields generally appear radiatively even if they are not
present in the initial Lagrangian. Some phenomenological consequences have
been also studied within the 5D orbifold theories in \cite{18,19}. Finally,
we should point out that some results of the recent papers \cite{19} and 
\cite{19a} partially overlap with the results of Section 3 of the present
work. Other localized operators have been also discussed in the literature
[21-24].

The paper is organized as follows. In Section 2 we will briefly remind the
basic facts about N=1 supersymmetry in 5D and the orbifold compactification
and set-up a simple realistic 5D SU(5) GUT. In Section 3 we discuss
Kaluza-Klein (KK) reduction of theory with kinetic terms localized on the
orbifold fixed points. In Section 4 we will discuss 1-loop contributions of
KK modes to the evolution of gauge couplings and the problem of gauge
coupling unification. The final Section 5 will be devoted to summary and
conclusions.

\section{Orbifold compactification and 5D SU(5) GUT}

We begin with a brief review of the orbifold compactification in the
simplest case of a only one extra dimension. The 5D space-time is a direct
product of 4D Minkowski space-time $M^{4}$ and an extra dimension
compactified on the orbifold $S^{1}/Z_{2},$ with coordinates $x^{M}$, where $%
M=0,1,2,3,5$ ($x^{5}\equiv y$). The $S^{1}/Z_{2}$ orbifold can be viewed as
a circle of radius $R$ with opposite points identified by action of $Z_{2}$
orbifold parity: $Z_{2}$: $y\rightarrow -y$. The actual physical space
therefore is the interval $y\in \left[ 0,\pi R\right] $ with two orbifold
fixed points at $y=0$ and $y=\pi R$. Under the $Z_{2}$ symmetry, a generic
5D bulk field $\phi (x^{\mu },y)$ ($\mu =0,1,2,3$) has a definite
transformation property 
\begin{equation}
\phi (x^{\mu },-y)=P\phi (x^{\mu },y),  \label{2.1}
\end{equation}
where the eigenvalues of $P$ must be $\pm 1$. Generally the field $\phi
(x^{\mu },y)$ also can have $U(1)$-twisted transformation under the periodic
translation along the fifth coordinate: $y\rightarrow y+2\pi R$: 
\begin{equation}
\phi (x^{\mu },y+2\pi R)=U\phi (x^{\mu },y),  \label{2.2}
\end{equation}
where $U_{\kappa }=\exp \left( 2i\pi \kappa \right) $. In fact this is the
well-known Hosotani-Scherk-Schwarz boundary condition. The case $\kappa =%
\frac{1}{2}$ is a discrete version of more general case (\ref{2.2}).
Denoting the field with $\left( P,U_{\frac{1}{2}}\right) =\left( \pm 1,\pm
1\right) $ by $\phi ^{(\pm ,\pm )}$, we obtain the following KK mode
expansion: 
\begin{eqnarray}
\phi ^{(+,+)}(x^{\mu },y) &=&\sum_{n=0}^{\infty }\sqrt{\frac{2}{2^{\delta
_{n,0}}a}}f_{n}^{(+,+)}(x^{\mu })\cos \left( \frac{ny}{R}\right) ,  \nonumber
\\
\phi ^{(+,-)}(x^{\mu },y) &=&\sum_{n=0}^{\infty }\sqrt{\frac{2}{a}}%
f_{n+1}^{(+,-)}(x^{\mu })\cos \left( \frac{\left( n+\frac{1}{2}\right) y}{R}%
\right) ,  \nonumber \\
\phi ^{(-,+)}(x^{\mu },y) &=&\sum_{n=0}^{\infty }\sqrt{\frac{2}{a}}%
f_{n+1}^{(-,+)}(x^{\mu })\sin \left( \frac{\left( n+1\right) y}{R}\right) , 
\nonumber \\
\phi ^{(-,-)}(x^{\mu },y) &=&\sum_{n=0}^{\infty }\sqrt{\frac{2}{a}}%
f_{n+1}^{(-,-)}(x^{\mu })\sin \left( \frac{\left( n+\frac{1}{2}\right) y}{R}%
\right) ,  \label{2.3}
\end{eqnarray}
where $a=\pi R$. Upon compactification the 4D fields (KK modes) $%
f_{n}^{(+,+)}(x^{\mu }),$ $f_{n+1}^{(+,-)}(x^{\mu }),$ $f_{n+1}^{(-,+)}(x^{%
\mu }),$ $f_{n+1}^{(-,-)}(x^{\mu })$ acquire masses $\frac{n}{R},$ $\frac{%
\left( n+\frac{1}{2}\right) }{R},$ $\frac{\left( n+1\right) }{R},$ $\frac{%
\left( n+\frac{1}{2}\right) }{R}$, respectively. Note, that zero mode is
contained only in $\phi ^{(+,+)}$ field. We will call (\ref{2.3}) the
standard KK decomposition. In next Section we will argue that this
decomposition is not valid in general.

Now let us describe the simplest N=1 supersymmetric SU(5) model in 5D. In
the 5D bulk we have SU(5) gauge supermultiplet, transforming as an adjoint
representation ${\cal V}^{A}\sim 24$ $(A=1,...,24)$ and two Higgs
hypermultiplets ${\cal H}$ and $\overline{{\cal H}}$ that transform as $5$
and $\overline{5}$, respectively. The 5D gauge supermultiplet contains a
vector boson $A_{M}^{A}$, two gauginos, $\lambda _{L}^{A}$ and $\lambda
_{R}^{A}$, and a real scalar $\sigma ^{A}$, which can be decomposed into a
vector supermultiplet $V^{A}=\left( A_{\mu }^{A},\lambda _{L}^{A}\right) $,
and a chiral supermultiplet $\Sigma ^{A}=((\sigma ^{A}+iA_{5}^{A})/\sqrt{2},$
$\lambda _{R}^{A})$ under 4D N=1 supersymmetry. The hypermultiplet ${\cal H}$
contains two complex scalars, $h$ and $h^{c},$ and two Weyl fermions, $\psi $
and $\psi ^{c}$. They can be combined into two 4D N=1 chiral supermultiplets 
$H=(h,\psi )$ and $H^{c}=(h^{c},\psi ^{c})$, which transform as $5$ and $%
\overline{5}$ under the SU(5) gauge group (similarly for $\overline{{\cal H}}
$).

\begin{table}[t] \centering%
$
\begin{tabular}{|c|c|c|c|}
\hline\hline
$(P,$ $U_{\frac{1}{2}})$ & Gauge and Higgs fields & Bulk matter fields & KK
masses \\ \hline\hline
$\left( +,+\right) $ & $V^{a}$, $H_{D}$, $\overline{H}_{D}$ & 10$_{U,E}$, 10$%
_{Q}^{\prime }$, $\overline{5}_{D}$, $\overline{5}_{L}^{\prime }$ & $\frac{n%
}{R}$ \\ \hline\hline
$\left( +,-\right) $ & $V^{\widehat{a}}$, $H_{T}$, $\overline{H}_{T}$ & 10$%
_{Q}$, 10$_{U,E}^{\prime }$, $\overline{5}_{L}$, $\overline{5}_{D}^{\prime }$
& $\frac{n+\frac{1}{2}}{R}$ \\ \hline\hline
$\left( -,+\right) $ & $\Sigma ^{a}$, $H_{D}^{c}$, $\overline{H}_{D}^{c}$ & 
10$_{U,E}^{c}$, 10$_{Q}^{\prime c}$, $\overline{5}_{D}^{c}$, $\overline{5}%
_{L}^{\prime c}$ & $\frac{n+1}{R}$ \\ \hline\hline
$\left( -,-\right) $ & $\Sigma ^{\widehat{a}}$, $H_{T}^{c}$, $\overline{H}%
_{T}^{c}$ & 10$_{Q}^{c}$, 10$_{U,E}^{\prime c}$, $\overline{5}_{L}^{c}$, $%
\overline{5}_{D}^{\prime c}$ & $\frac{n+\frac{1}{2}}{R}$ \\ \hline\hline
\end{tabular}
$%
\caption{The particle content (bulk fields) of the orbifold SU(5) GUT, their transformation 
properties under the orbifold symmetries and KK masses according to the standard decomposition (\ref{2.3}). }%
\end{table}%

The 5D SU(5) symmetry is broken down to the to the SU(3)$\otimes $SU(2)$%
\otimes $U(1) Standard Model group by orbifold boundary conditions. This can
be achieved in different ways. We will assume that the gauge symmetry
breaking occurs because of non-trivial periodic boundary conditions. Namely,
we choose: $U_{\frac{1}{2}}=(-1,-1,-1,+1,+1)$ and $P=(+1,+1,+1,+1,+1)$,
where $U_{\frac{1}{2}}$ and $P$ act on the fundamental representation of
SU(5). At the same time we have to assign opposite $Z_{2}$ orbifold parities
to the 4D N=2 partners of the bulk fields, in order to keep only N=1
supersymmetry in 4D effective theory. The orbifold symmetries for all
components of the vector and Higgs multiplets are shown in Table 1. Here, we
split SU(5) index $A=a,\widehat{a}$ into the indices, $a$ and $\widehat{a},$
which correspond to the unbroken and broken SU(5) generators, respectively.
The subscripts $T$ and $D$ denote colour triplet and weak doublet components
of the Higgs multiplets, respectively. Since only $(+,+)$ fields have zero
modes, the massless sector of the model consists of SU(3)$\otimes $SU(2)$%
\otimes $U(1) N=1 vector multiplet $V_{0}^{a}$ and Higgs doublet and
anti-doublet chiral superfields $H_{D}$ and $\overline{H}_{D}$, colour
triplet chiral superfields $H_{T}$ and $\overline{H}_{T},$ are massive, thus
realizing doublet-triplet splitting in a simple geometric way. This also
means that Higgsino mediated proton decay is absent in the model. Non-zero
modes of at each KK level $n$ $V^{a}$ eat corresponding $\Sigma ^{a} $
becoming massive and similarly for the X-Y vector multiplets $V^{\widehat{a}%
} $ and $\Sigma ^{\widehat{a}}$. Note also that locally at $y=0$ fixed point
SU(5) symmetry is exact, while it is broken on the $y=\pi R$ fixed point.
Thus the two fixed points are not equivalent.

Each generation of matter fields, are placed into hypermultiplets which
transform, as usually, as $\overline{5}$ and $10$ representations of SU(5).
There are different ways how to put matter fields in the bulk. They can be
localized on one of the fixed points. In order to preserve the successful
SU(5) $b$-$\tau $ unification it is desirable to have third generation
matter on the SU(5)-preserving fixed point at $y=0$. On the other hand, one
can place matter fields in the bulk as well. An important point is that, in
such a case one has to double the representation $\overline{5}+10$ by
introducing $\overline{5}^{\prime }+10^{\prime }$ \cite{5}. One can see from
Table 1, that zero-mode matter fields, i.e. ordinary quarks and leptons
(subscripts $Q,L,U,D$ and $E$ denote quark doublet, lepton doublet,
up-antiquark, down-antiquark and positron components of the corresponding
SU(5) representations, respectively), come now from different
representations. This means that X-Y gauge boson can not be responsible for
the proton decay in case of all three generations being in the bulk. The
proton decay will be significantly suppressed if only light generations (or
part of them) are residing in the bulk, while third generation is localized
on SU(5)-symmetric fixed point \cite{6,7}.

The model described above is indeed very attractive. One can see, that most
of the difficulties of the ordinary 4D SU(5) GUT can be resolved in a very
simple way. It remains to see whether one can have in the above picture
reliable predictions. But before discussing gauge coupling unification in
the above model we would like to come back to the question of KK mode
decomposition in orbifold field theories.

\section{KK decomposition in 5D with orbifold compactification}

Once again, we are considering 5D space-time with the fifth extra dimension $%
y$ being an $S^{1}/Z_{2}$ orbifold, i.e. a line of length $a=\pi R$ with two
fixed points at $y=0$ and $y=a$. As we have mentioned in the Introduction,
generally one is allowed to add to the 5D bulk Lagrangian ${\cal L}_{5}$
Lagrangians ${\cal L}_{0}$ and ${\cal L}_{\pi }$ localized at $y=0$ and $y=a$
4D subspaces. So the total Lagrangian is: 
\begin{equation}
{\cal L}={\cal L}_{5}+\delta (y){\cal L}_{0}+\delta (y-a){\cal L}_{\pi }.
\label{3.1}
\end{equation}
${\cal L}_{0}$ and ${\cal L}_{\pi }$ could contain various operators which
respect the symmetries on the fixed points but not in the full 5D bulk. In
certain cases of interest, however, they might be avoided by some symmetries
(supersymmetry). In realistic models this is not the case for the kinetic
terms. Indeed, the localized kinetic terms for certain 5D bulk fields
inevitably appear (unless one is considering phenomenologically unacceptable
theory with conformal invariance) as a result of radiative corrections.
Contributions to the radiative induced local kinetic terms come not only
from the fields localized at the fixed points but from the bulk fields as
well \cite{17}. These contributions are logarithmically divergent and
consistency of the theory requires introduction of the corresponding local
counter-terms. Residual finite parts (after cancellation of divergences) are
not calculable within the effective orbifold field theory and thus should be
treated as new free parameters. Basically, they can be calculated only
within a more general theory which substitutes the low energy effective
orbifold theory at some ultraviolet scale $\Lambda $. Thus, the free
parameters of the local kinetic terms parametrize our ignorance of the
fundamental UV physics.

In turn, the local kinetic terms could significantly affect the entire
phenomenology of the effective low energy orbifold theories. Namely the KK
mode decomposition discussed in Section 2 is not valid anymore. To see this,
let us consider the simple case of 5D massless scalar field: 
\begin{eqnarray}
{\cal L}_{5} &=&\partial _{M}\Phi ^{+}\partial ^{M}\Phi ,  \nonumber \\
{\cal L}_{0} &=&r_{0}\partial _{\mu }\Phi ^{+}\partial ^{\mu }\Phi , 
\nonumber \\
{\cal L}_{\pi } &=&r_{\pi }\partial _{\mu }\Phi ^{+}\partial ^{\mu }\Phi ,
\label{3.2}
\end{eqnarray}
where $r_{0}$ and $r_{\pi }$ are new scale parameters associated with local
kinetic terms at $y=0$ and $y=a$, respectively. The parameters $r_{0}$
and/or $r_{\pi }$ are vanishing for the field $\Phi $ which vanishes at $y=0$
and/or $y=a$ fixed points. The behaviour of $\Phi $ at fixed points in turn
is defined by the orbifold symmetries at hand. We consider general case
where $\Phi $ is either even or odd under the orbifold parity $Z_{2}$ (\ref
{2.1}) and it is $U(1)$-twisted under the periodic translation (\ref{2.2}).
The equation of motion followed from (\ref{3.1}) and (\ref{3.2}) is: 
\begin{equation}
\left( \left[ 1+r_{0}\delta (y)+r_{\pi }\delta (y-a)\right] \partial _{\mu
}\partial ^{\mu }-\partial _{5}^{2}\right) \Phi (x^{\mu },y)=0.  \label{3.5}
\end{equation}
It is supplemented by the boundary conditions (\ref{2.1}) and (\ref{2.2}).
Performing the KK decomposition of $\Phi $: 
\begin{equation}
\Phi (x^{\mu },y)=\sum_{m_{n}}\phi _{m_{n}}(x^{\mu })f_{m_{n}}(y),\text{ }%
-\infty \leq m_{n}\leq +\infty   \label{3.6}
\end{equation}
the Eq. (\ref{3.5}) is split into the Klein-Gordon equation 
\begin{equation}
\left( \partial _{\mu }\partial ^{\mu }+m_{n}^{2}\right) \phi
_{m_{n}}(x^{\mu })=0  \label{3.7}
\end{equation}
for the modes $\phi _{n}(x^{\mu })$ with KK masses $m_{n}$ and the equation
which determines the $y$ dependence of the modes: 
\begin{equation}
\left( \partial _{5}^{2}+m_{n}^{2}\left[ 1+r_{0}\delta (y)+r_{\pi }\delta
(y-a)\right] \right) f_{m_{n}}(y)=0.  \label{3.8}
\end{equation}
Eq. (\ref{3.8}) can be solved by the method of images. The problem is
actually reduced to a simple one-dimensional quantum-mechanical problem of a
particle moving in a periodic potential formed by a sequence of Dirac $%
\delta $-functions at $y=0+2an$ and $y=a+2an$. Since the equation of motion
and the boundary conditions are periodic, we can apply Floquet's theorem and
write down a general solution in different regions as: 
\begin{equation}
f_{m_{n}}(y)=\left\{ 
\begin{array}{c}
A_{m_{n}}e^{im_{n}y}+B_{m_{n}}e^{-im_{n}y}, \\ 
\\ 
C_{m_{n}}e^{im_{n}y}+D_{m_{n}}e^{-im_{n}y}, \\ 
\\ 
e^{i2a\kappa }\left[ A_{m_{n}}e^{im_{n}y}+B_{m_{n}}e^{-im_{n}y}\right] , \\ 
\\ 
e^{i2a\kappa }\left[ C_{m_{n}}e^{im_{n}y}+D_{m_{n}}e^{-im_{n}y}\right] ,
\end{array}
\begin{array}{c}
0<y<a \\ 
\\ 
a<y<2a \\ 
\\ 
-a<y<0 \\ 
\\ 
-2a<y<-a
\end{array}
\right. ,  \label{3.9}
\end{equation}
where $A_{m_{n}},B_{m_{n}},C_{m_{n}},D_{m_{n}}$ are constants to be
determined by matching the function $f_{n}(y)$ and its first derivatives at
fixed points. Assuming that $f_{n}(y)$ is continuous across the fixed
points, we get the following system of homogeneous algebraic equations for $%
A_{m_{n}},B_{m_{n}},C_{m_{n}},D_{m_{n}}$: 
\begin{equation}
\widehat{{\cal M}}_{n}\left[ 
\begin{array}{c}
A_{m_{n}} \\ 
B_{m_{n}} \\ 
C_{m_{n}} \\ 
D_{m_{n}}
\end{array}
\right] =0,  \label{3.10}
\end{equation}
where 
\begin{equation}
\widehat{{\cal M}}_{n}=\left[ 
\begin{array}{llll}
1, & 1, & -e^{i2a\left( \kappa +m_{n}\right) }, & -e^{i2a\left( \kappa
-m_{n}\right) } \\ 
e^{-iam_{n}}, & e^{-iam_{n}}, & -e^{-iam_{n}}, & -e^{-iam_{n}} \\ 
\left( r_{0}m_{n}^{2}+2im_{n}\right)  & r_{0}m_{n}^{2} & -2im_{n}e^{i2a%
\left( \kappa +m_{n}\right) }, & 0 \\ 
\left( r_{\pi }m_{n}^{2}-2im_{n}\right) e^{iam_{n}}, & r_{\pi
}m_{n}^{2}e^{-iam_{n}}, & 2im_{n}e^{iam_{n}}, & 0
\end{array}
\right] .  \label{3.11}
\end{equation}
The necessary and sufficient condition for the existence of a nontrivial
solution of (\ref{3.10}), 
\begin{equation}
\det \widehat{{\cal M}}_{n}=0,  \label{3.12}
\end{equation}
is actually nothing but the KK mass quantization condition.

Solving the system (\ref{3.10}) one determines three out of four constants $%
A_{m_{n}},B_{m_{n}},C_{m_{n}},D_{m_{n}}$. The remaining one is determined
from the normalization of $f_{m_{n}}(y)$. In the presence of local kinetic
terms the wave functions $f_{m_{n}}(y)$ are generally not orthogonal, $%
\int_{0}^{a}dy$ $f_{m_{n}}^{*}(y)^{*}f_{m_{k}}(y)\neq \delta _{m_{n}m_{k}}$.
This would imply that in the interactions involving these modes the KK
number is not conserved, as it was obviously expected from the beginning,
since the translational invariance along the fifth dimension is explicitly
broken. In order to get the canonical kinetic terms for the KK modes upon
the integration over the extra dimension, we imply the following
orthonormality condition: 
\begin{equation}
\int_{0}^{a}dy\left[ 1+r_{0}\delta (y)+r_{\pi }\delta (y-a)\right]
f_{m_{n}}^{*}(y)f_{m_{k}}(y)=\delta _{m_{n}m_{k}}.  \label{3.13}
\end{equation}
Note that the $\left[ 1+r_{0}\delta (y)+r_{\pi }\delta (y-a)\right] $ acts
as a nontrivial volume-factor in the integral over $y$. Obviously, if the
field $\Phi (x^{\mu },y)$ is real, i.e. $\phi _{m_{n}}(x^{\mu })=\phi
_{m_{n}}^{+}(x^{\mu }),$ $f_{m_{n}}^{*}(y)=f_{-m_{n}}(y)$, one can recast
the decomposition (\ref{3.6}) into the form: 
\begin{equation}
\Phi (x^{\mu },y)=\sum_{m_{n}\geq 0}\phi _{m_{n}}(x^{\mu })\left(
f_{m_{n}}(y)+f_{-m_{n}}(y)\right) .\text{ }  \label{3.14}
\end{equation}
Then the normalization constants get multiplied by a factor of 2 relative to
the case of complex field.

Now let us consider some particular examples of fields with different
orbifold symmetries.

\paragraph{Z$_{2}$--even, periodic field, $\Phi ^{(+,+)}(x^{\mu },y)$.}

Since the Z$_{2}$-even field is nonvanishing at both boundaries, $r_{0}$ and 
$r_{\pi }$ are in general non-zero. Taking $\kappa =0$ (periodic boundary
condition), we obtain the following solution of the Eqs. (\ref{3.10}): 
\begin{equation}
f_{m_{n}}(y)=A_{m_{n}}\left( e^{im_{n}y}+\frac{2im_{n}+r_{0}m_{n}^{2}}{%
2im_{n}-r_{0}m_{n}^{2}}e^{-im_{n}y}\right) ,\text{ }0\leq y\leq a.
\label{3.15}
\end{equation}
Then the KK mass quantization condition (\ref{3.12}) becomes: 
\begin{equation}
\tan \left( m_{n}a\right) =-\frac{m_{n}\left( r_{0}+r_{\pi }\right) }{%
2\left( 1-\frac{r_{0}r_{\pi }m_{n}^{2}}{4}\right) }  \label{3.16}
\end{equation}
and the normalization constants $A_{m_{n}}$ are: 
\begin{equation}
A_{0}=\frac{1}{\sqrt{2\left( 2a+r_{0}+r_{\pi }\right) }}  \label{3.17}
\end{equation}
for the massless zero mode ($m_{0}=0$) and 
\begin{equation}
A_{m_{n}}=\frac{1}{\sqrt{2a+\frac{\left( r_{0}+r_{\pi }\right) \left( 1+%
\frac{r_{0}r_{\pi }m_{n}^{2}}{4}\right) }{\left( 1+\frac{r_{0}^{2}m_{n}^{2}}{%
4}\right) \left( 1+\frac{r_{\pi }^{2}m_{n}^{2}}{4}\right) }}}  \label{3.18}
\end{equation}
for non-zero modes ($m_{n}\neq 0$).

The Eq. (\ref{3.16}) for KK masses $m_{n}$ can be solved numerically. Here
we consider some analytic approximations assuming first that all $r$-factors
are positive. If $r$-factors are small enough, $\frac{r_{0,\pi }}{a}\ll 1,$
KK mode decomposition is reduced to the standard one (see (\ref{2.3})) with
KK masses $m_{n}\approx \frac{n}{R},$ $n=0,1,...$ However, for large $r$%
-factors, $\frac{r_{0,\pi }}{a}\gtrsim 1,$ KK modes substantially deviate
from those discussed in the previous Section. As an extreme case consider $%
\frac{r_{0,\pi }}{a}\gg 1$ ($r_{0}\approx r_{\pi }\equiv r,$ $\xi =\frac{r}{a%
}$)$.$ Then KK modes at each level $n\geq 2$ once again approach their
standard values 
\begin{equation}
m_{n}\approx \frac{n}{R}-\frac{4}{\xi ^{2}a\pi n},\text{ }n=2,3...
\label{3.18a}
\end{equation}
while the first excited mode is extremely light: 
\begin{equation}
m_{1}\approx \frac{2}{a}\sqrt{\frac{1-\frac{1}{\xi }}{\xi }}\approx \frac{2}{%
\pi R\sqrt{\xi }}<<\frac{1}{R}.  \label{3.18b}
\end{equation}
The appearance of the light mode in this limit is a peculiar property of the
presence of two opaque boundaries \cite{19}. Nothing similar happens if one
sets one of the two $r$-factors to zero. In the case of graviton (instead of
the scalar field) with such sufficiently light mode one could have, e.g., an
interesting new bigravity model \cite{24}.

\paragraph{Z$_{2}$--even, anti-periodic field, $\Phi ^{(+,-)}(x^{\mu },y)$.}

In this case the field vanishes on the boundary $y=a$. Thus taking $r_{\pi
}=0$ and $\kappa =\frac{1}{2}$ (anti-periodic boundary condition), we obtain
the solution 
\begin{equation}
f_{m_{n}}(y)=A_{m_{n}}\left( e^{im_{n}y}-e^{-im_{n}\left( y-2a\right)
}\right) ,\text{ }0\leq y\leq a,  \label{3.19}
\end{equation}
which is amended by the quantization condition 
\begin{equation}
\cot \left( m_{n}a\right) =\frac{r_{0}m_{n}}{2}.  \label{3.20}
\end{equation}
The normalization constants in (\ref{3.19}) are 
\begin{equation}
A_{m_{n}}=\frac{1}{\sqrt{2a+\frac{r_{0}}{1+\frac{r_{0}^{2}m_{n}^{2}}{4}}}}
\label{3.21}
\end{equation}
Once more one can discuss analytically some limiting approximations. For
small $r_{0},\frac{r_{0}}{a}\equiv \xi \ll 1$ we essentially reproduce the
standard KK modes and their masses (\ref{2.3}), 
\begin{equation}
m_{n}\approx \frac{(n+\frac{1}{2})}{R\left( 1+\frac{\xi }{2}\right) }.
\label{3.21a}
\end{equation}
An opposite limit of large $\xi \gg 1$ gives KK masses substantially lighter
compared to the standard case: 
\begin{equation}
m_{n}\approx \frac{n}{R}+\frac{2}{\xi a\pi n}<\frac{(n+\frac{1}{2})}{R},
\label{3.21b}
\end{equation}
for $n=2,3,...$ and a lighter first excited mode: 
\begin{equation}
m_{1}\approx \sqrt{\frac{2}{\xi }}\frac{1}{\pi R}<<\frac{1}{2R}.
\label{3.21c}
\end{equation}

\paragraph{Z$_{2}$--odd, anti-periodic field, $\Phi ^{(-,-)}(x^{\mu },y)$.}

In this case the boundary at $y=0$ is transparent for the KK modes. Thus
taking $r_{0}=0$ and $\kappa =\frac{1}{2}$ we obtain 
\begin{equation}
f_{m_{n}}(y)=A_{m_{n}}\left( e^{im_{n}y}+e^{-im_{n}\left( y-2a\right)
}\right) ,\text{ }0\leq y\leq a.  \label{3.22}
\end{equation}
Mass quantization condition and normalization constants can be obtained
changing $r_{0}$ by $r_{\pi }$ in the corresponding formulas (\ref{3.20})
and (\ref{3.21}). Thus in the case of equivalent fixed points ($r_{0}=r_{\pi
}$) KK masses of $\left( +,-\right) $ and $(-,-)$ modes are degenerate. The
approximations of small and large $r_{\pi }$ are evidently given by the same
formulas (\ref{3.21a}) and (\ref{3.21b}), (\ref{3.21c}) (with $\xi =\frac{%
r_{\pi }}{a}$), respectively.

\paragraph{{\bf Z}$_{2}$--odd, periodic field, $\Phi ^{(-,+)}(x^{\mu },y)$,}

is unaffected by the presence of boundaries, since orbifold symmetries force
the field to be vanishing at both boundaries. Hence the standard KK mode
decomposition discussed in Section 2 remains unchanged in this case.

Before going further, few remarks concerning $r$-factors are in order. So
far we have assumed that $r_{0}$ and $r_{\pi }$ are positive. Treating them
as free parameters one can ask whether both or one of them can be negative.
That is to say, whether the local kinetic terms on the fixed points can have
wrong sign. An obvious constraint in such cases comes from the appearance of
negative norm states. One can see, however, that in certain cases the ghost
KK modes can be avoided. Consider for example $(+,+)$ modes and assume $%
r_{0}=-r_{\pi }$. Then, from (\ref{3.16}), (\ref{3.17}) and (\ref{3.18}), it
is clear that the KK modes behave as in the standard case without local
kinetic terms, whenever $r$-factors are large or small. The contributions
from the localized kinetic terms are cancelled out and boundaries become
transparent for the corresponding modes. For the $(+,-)$ ($(-,-)$) modes and
negative $r_{0}$ ($r_{\pi }$), ghost-free effective theory would imply $%
2a\left( 1+\frac{r_{0}^{2}m_{n}^{2}}{4}\right) +r_{0}>0$ ($2a\left( 1+\frac{%
r_{\pi }^{2}m_{n}^{2}}{4}\right) +r_{\pi }>0$) for all $n$. Obviously, the
above ghost-free condition can be satisfied, for instance, when $r_{0}>-2a$ (%
$r_{\pi }>-2a$). Many other possibilities can be found as well. The overall
massage here is that the KK mass spectrum and wave functions crucially
depend not only on the size of $r$-factors but on their signs as well.

Another important point to stress is that $r$-factors (in general case when
interactions are included) are actually scale dependent and undergo
renormalization. Thus the physical KK masses should be determined using the
corresponding renormalization group equations. Note, however, that in
certain cases of interest this effects might be negligible.

Having determined the KK masses and wave functions from the free
(linearized) theory one can discuss now the interactions. As a
representative example let us add to the free Lagrangian (\ref{3.1}), (\ref
{3.2}), a $\varphi ^{4}$-potential ${\cal V}$: 
\begin{equation}
{\cal V}=\left[ \lambda _{5}+\lambda _{0}\delta (y)+\lambda _{\pi }\delta
(y-a)\right] \left( \Phi ^{+}\Phi \right) ^{2},  \label{3.23}
\end{equation}
where $\lambda _{5}$ is a 5D coupling, while $\lambda _{0}$ and $\lambda
_{\pi }$ are couplings of the interaction terms localized at boundaries $y=0$
and $y=a$, respectively. The effective 4D couplings can be derived upon KK
mode expansion. They are given as: 
\begin{equation}
\lambda _{ijkl}=\lambda _{5}\int_{0}^{a}dy\left[ 1+\frac{\lambda _{0}}{%
\lambda _{5}}\delta (y)+\frac{\lambda _{\pi }}{\lambda _{5}}\delta
(y-a)\right] f_{m_{i}}^{*}f_{m_{j}}f_{m_{k}}^{*}f_{m_{l}}.  \label{3.24}
\end{equation}
Note, that the KK number is violated in these interactions in general even
if the localized potentials are absent, $\lambda _{0},\lambda _{\pi }=0$.
Only in the standard case when the local kinetic terms are also absent the
KK-number (i.e. the fifth component of the momentum) is conserved.

The effective 4D couplings (\ref{3.24}) are determined through the rather
complicated integrals of the product of wave functions $f_{m_{n}}.$ However,
one can readily say something about interactions involving zero modes.
Consider for instance $\lambda _{00kl}$. Once again in general KK-number is
still not conserved. Suppose now that $\frac{\lambda _{0}}{\lambda _{5}}%
=r_{0}$ and $\frac{\lambda _{\pi }}{\lambda _{5}}=r_{\pi }$. Then (\ref{3.24}%
) is considerably simplified and one gets (using orthonormality condition (%
\ref{3.13})): 
\begin{equation}
\lambda _{00kl}=\left| A_{0}\right| ^{2}\lambda _{5}\delta _{kl}.
\label{3.25}
\end{equation}
These interactions conserve the KK-number and are universal for all KK
modes. The effective 4D coupling is given by: 
\begin{equation}
\lambda _{4}=\frac{\lambda _{5}}{2\left( 2a+r_{0}+r_{\pi }\right) }.
\label{3.26}
\end{equation}
This is the familiar relation between 4D and 5D couplings except the fact
that the ''volume'' of extra space, $a,$ is replaced by the effective
volume, $a+r_{0}/2+r_{\pi }/2$. So, if the bulk and localized couplings are
aligned with kinetic $r$-factors in the way assumed above, the interaction
of zero modes with those of non-zero KK modes are universal and given by (%
\ref{3.26}). This is obviously true also when zero modes interact with
different bulk fields with different orbifold symmetries, provided, of
course, that the interactions respect these symmetries as it is required by
the consistency of the theory.

In fact the above situation is exactly what happens in the case of bulk
gauge theories. Indeed, consider some non-Abelian gauge field in the bulk
with local kinetic terms on the boundaries: 
\begin{equation}
{\cal L}=-\frac{1}{4}F_{MN}^{a}F^{aMN}-\frac{1}{4}\frac{g_{5}^{2}}{g_{0}^{2}}%
\delta (y)F_{\mu \nu }^{a}F^{a\mu \nu }-\frac{1}{4}\frac{g_{5}^{2}}{g_{\pi
}^{2}}\delta (y-a)F_{\mu \nu }^{a}F^{a\mu \nu },  \label{3.27}
\end{equation}
where $F_{MN}^{a}=\partial _{M}A_{N}^{a}-\partial
_{N}A_{M}^{a}+g_{5}f^{abc}A_{M}^{b}A_{N}^{c}$ is the gauge field strength, $%
g_{5}$ bulk gauge coupling and $g_{0}$ and $g_{\pi }$ couplings for the
localized terms. In this case $r$-factors are given by $r_{0}=\frac{g_{5}^{2}%
}{g_{0}^{2}}$ and $r_{\pi }=\frac{g_{5}^{2}}{g_{\pi }^{2}}$. Linearized
equations for the wave functions $f_{m_{n}}$ (they are real in this case)
have exactly the same form (in the unitary gauge, $A_{5}=0$) as in the case
of scalar field discussed above (see Eq. (\ref{3.8})). The massless zero
modes of gauge field ($A_{\mu }^{a}$ assumed to be Z$_{2}$-even and
periodic) have the following self-interactions: 
\begin{eqnarray}
g_{0kl} &=&\frac{g_{5}}{\sqrt{a+\frac{r_{0}}{2}+\frac{r_{\pi }}{2}}}\delta
_{kl}\equiv g_{4},  \nonumber \\
g_{00kl} &=&\frac{g_{5}^{2}}{a+\frac{r_{0}}{2}+\frac{r_{\pi }}{2}}\delta
_{kl}\equiv g_{4}^{2}.  \label{3.28}
\end{eqnarray}
The zero mode also universally interacts with bulk matter fields
irrespective of the $r$-factors of the local kinetic terms of the matter
fields (which might be different from those for the gauge field) and matter
fields localized on the orbifold fixed points. Clearly, all these features
are dictated by the gauge invariance. Note, however, that the above
statement is not true for the KK modes of the gauge fields. For instance,
fields localized on different fixed points would interact with the KK mode
of the same gauge boson differently, if the fixed points are not equivalent.

From the above discussion it is clear that the localized kinetic terms will
have profound consequences for the entire phenomenology of
higher-dimensional theories. Modification of collider phenomenology in some
limited cases has been recently studied in \cite{16,18,19}. In the following
Section we will discuss the effects of localized kinetic terms on the gauge
coupling unification in the framework of 5D GUT model described in Section 2.

\section{Gauge coupling unification in realistic SU(5) orbifold GUT}

An important consistency check of any GUT model is its prediction for the
standard gauge couplings $\alpha _{1}(M_{Z}),$ $\alpha _{2}(M_{Z}),$ $\alpha
_{3}(M_{Z})$ at low energies, namely at $M_{Z}$ (Z-boson mass). They are
related to the unified gauge coupling $\alpha _{GUT}(M_{GUT})\equiv \alpha
_{1}(M_{GUT})=$ $\alpha _{2}(M_{GUT})=$ $\alpha _{3}(M_{GUT})$ at the
unification scale $M_{GUT\text{ }}$through the renormalization group
equations. At two-loop level the minimal 4D supersymmetric SU(5) model gives
the following predictions for the strong gauge coupling $\alpha _{3}(M_{Z})$%
: 
\begin{equation}
\alpha _{3}^{SU(5)}(M_{Z})\simeq 0.130\pm 0.004\pm \Delta ^{SU(5)}
\label{4.1}
\end{equation}
and for the unification scale: 
\begin{equation}
M_{GUT}^{SU(5)}\simeq 2\cdot 10^{16}\text{ GeV,}  \label{4.2}
\end{equation}
where the experimental values for $\alpha _{1}(M_{Z})$ and $\alpha
_{2}(M_{Z})$ couplings are taken as input parameters. The prediction (\ref
{4.1}) should be compared with the high-precision experimental value \cite
{25}, 
\begin{equation}
\alpha _{3}^{\exp }(M_{GUT})=0.119\pm 0.002.  \label{4.3}
\end{equation}
The main uncertainty in (\ref{4.1}), $\Delta ^{SU(5)}$, originates from the
uncertainties in masses of GUT particles, while the first uncertainty is
related to the variations in the mass spectrum of the electroweak Higgs
bosons and superparticles at the TeV-scale. In the minimal supersymmetric
SU(5), $\Delta ^{SU(5)}$ is parametrized through the coloured Higgs
(Higgsino) mass $M_{T}$, $\Delta ^{SU(5)}=\frac{3\alpha _{GUT}}{10\pi }\ln
\left( \frac{M_{T}}{M_{GUT}}\right) $. Thus, large negative contribution $%
\Delta ^{SU(5)}\approx -0.01$ needed to fit the experimental value (\ref{4.1}%
) can be obtained by lowering the mass $M_{T}$, $M_{T}\sim 0.08M_{GUT}$.
However, this possibility is excluded from the unacceptably fast proton
decay induced by a low mass coloured Higgsino\footnote{%
Several mechanisms to suppress Higgsino mediated proton decay is known (see
e.g. \cite{26}). However, all of them require introduction of new particles
and/or interactions, and hence new parameters, beyond the minimal SU(5).
This, in turn, significantly reduces the predictive power of such models.} 
\cite{27}. Thus the minimal 4D SU(5) GUT indeed contradicts the data. In
more involved GUT models, the GUT threshold corrections depend on a larger
number of unknown parameters which can neither be constrained from
independent data, nor unambiguously predicted. In this situation we
essentially loose the predictivity.

In this respect, the situation in orbifold GUT, where GUT symmetry breaking
is achieved by orbifold compactification, seems more promising. Firstly,
ignoring for a moment possible operators localized on the orbifold fixed
points, the mass spectrum of GUT particles and their KK modes are
essentially determined by few parameters, namely in 5D by a single
compactification radius $R$. Secondly, one can avoid the proton decay
constraints by intrinsically geometric mechanism without extension of
particle content of the model. Thirdly, although each gauge coupling above
the compactification scale receives radiative corrections which depend on a
certain power of the cut-off scale, the relative slope of the gauge
couplings and thus the low energy predictions will have the usual
logarithmic scale dependence. This is because of the underlying bulk GUT
symmetry. However, as we have seen in the previous Section the local kinetic
terms might substantially disturb the KK mass spectrum and one may worry
whether one can actually retain the predictivity power.

In fact the problem appears already at tree level. Consider for definiteness
a realistic SU(5) GUT in 5D described in Section 2. Since on the orbifold
fixed point $y=a$ SU(5) symmetry is broken down to the SU(3)$\otimes $SU(2)$%
\otimes $U(1), one is allowed to write down only a SU(3)$\otimes $SU(2)$%
\otimes $U(1)-symmetric (rather than SU(5)-symmetric) gauge kinetic terms
with different $r$-factors, $r_{\pi }^{i}$ ($i=1,2,3$ correspond to U(1),
SU(2) and SU(3), respectively). Then the 4D effective gauge coupling (see
Eq. (\ref{3.28})) at some unification scale will be 
\begin{equation}
g_{4}^{i}(M_{GUT})=\frac{g_{5}(M_{GUT})}{\sqrt{a+\frac{r_{0}(M_{GUT})}{2}+%
\frac{r_{\pi }^{i}(M_{GUT})}{2}}}.  \label{4.4}
\end{equation}
If $r_{\pi }^{i}(M_{GUT})$ are large and non-universal, $r_{\pi
}^{i}(M_{GUT})\gtrsim a$, then the gauge coupling unification is ruined
already at the tree level. Some extra assumption beyond the SU(5) framework
is evidently needed to suppress the localized gauge kinetic terms at
SU(5)-breaking orbifold fixed point, i.e. $\frac{r_{\pi }^{i}}{a}$ should be 
$<<1$.

The explanation to why the $r$-factors of gauge localized kinetic terms
might be small has been proposed by Hall and Nomura \cite{6,7}. The key
assumption is that the theory enters into the strong gauge coupling regime
at $M_{GUT}$. This could naturally happen because above the compactification
scale the asymptotic freedom is lost and couplings grow rapidly due to the
power-low running of couplings in higher dimensions. Naive dimensional
analysis gives the following estimates for the strong bulk gauge coupling, $%
g_{5}(M_{GUT})\simeq \sqrt{\frac{16\pi ^{3}}{CM_{GUT}}},$ and for strong
gauge couplings for the local kinetic terms one has: $g_{0}(M_{GUT})\simeq 
\sqrt{\frac{16\pi ^{2}}{C}}$ and $g_{\pi }^{i}(M_{GUT})\simeq \sqrt{\frac{%
16\pi ^{2}}{C^{i}}}$ (here $C^{\prime }$s are group theoretic factors, $%
C\simeq 5$ for SU(5) and $C^{i}\simeq (1,2,3)$) \cite{7}. Within this
assumption $r$-factors are indeed small, $\frac{r_{0}(M_{GUT})}{a}=$ $\frac{%
g_{5}^{2}}{ag_{0}^{2}}<<1,$ $\frac{r_{\pi }^{i}(M_{GUT})}{a}=$ $\frac{%
g_{5}^{2}(M_{GUT})}{ag_{\pi }^{i2}(M_{GUT})}<<1,$ providing that extra
dimension is large, $aM_{GUT}>>1$ ($a=\pi R$). Since the low energy values
for the effective 4D couplings $g_{4}^{i}$ are known to be of the order $%
\sim 0.7,$ from the above assumption one can estimate the energy gap between
the compactification scale $\frac{1}{R}$ and strong coupling unification
scale $M_{GUT}$: $M_{GUT}R\simeq 60$. That is to say, there are
approximately $N_{KK}\simeq 60$ KK modes in the energy region between the
compactification scale $\frac{1}{R}$ and the unification scale $M_{GUT}$.

To see how this works, let us calculate the KK mode corrections in the model
described in the Section 2 with standard KK masses, see Table 1. One obtains
at one loop leading-log approximation: 
\begin{equation}
\alpha _{3}^{-1}(M_{Z})=\alpha _{3}^{-1SU(5)}(M_{Z})+\frac{6}{7\pi }%
\sum_{n=0}^{N_{KK}-1}\ln \left( \frac{n+1}{n+\frac{1}{2}}\right) ,
\label{4.4a}
\end{equation}
(here $\alpha _{3}^{-1SU(5)}(M_{Z})$ is the prediction of the 4D
supersymmetric SU(5) for (\ref{4.1}) without GUT thresholds, $\Delta
^{SU(5)}=0$) for the strong gauge coupling and 
\begin{equation}
\ln \left( \frac{M_{c}}{M_{Z}}\right) =\ln \left( \frac{M_{GUT}^{SU(5)}}{%
M_{Z}}\right) +\frac{4}{7}\sum_{n=0}^{N_{KK}-1}\ln \left( \frac{n+1}{n+\frac{%
1}{2}}\right) -\ln N_{KK}.  \label{4.4b}
\end{equation}
Note that the KK mode correction to $\alpha _{3}(M_{Z})$ (the last term in
Eq. (\ref{4.4a})) works in the right direction, lowering $\alpha
_{3}^{SU(5)}(M_{Z})$ (\ref{4.1}). Taking $N_{KK}=60$, one obtains correction
from KK modes $\delta \alpha _{3}^{-1}(M_{Z})\approx 0.715$ which to the
brings prediction for the strong gauge coupling, 
\begin{equation}
\alpha _{3}(M_{Z})\approx 0.119\pm 0.004,  \label{4.4c}
\end{equation}
in remarkable agreement with the experimental value (\ref{4.3}). At the same
time, the compactification scale $M_{c\text{ }}$is lower, 
\begin{equation}
M_{c\text{ }}\approx 1.5\cdot 10^{15}\text{ GeV,}  \label{4.4d}
\end{equation}
than the typical 4D unification scale (\ref{4.2}). This means that the first
KK modes of X-Y intermediate vector bosons might induce unacceptably fast
decay of proton. This inevitably forces us to put all or part of light
generations into the 5D bulk. Then the fast proton decay can be avoided \cite
{5,6,7}.

Let us now calculate possible corrections to the above result (\ref{4.4c}), 
\ref{4.4d}) due to the modification of the KK mass spectrum by the local
kinetic terms. Consider first gauge vector multiplets $V^{a}$ and $V^{%
\widehat{a}}$. The $r$-factors for these superfields are controlled by the
strong coupling unification assumption and has been estimated above. Thus,
the corresponding leading-log correction can be unambiguously computed. We
find: 
\begin{eqnarray}
\delta \alpha _{3}^{-1}(M_{Z}) &\approx &-\frac{N_{KK}}{2\pi }\times 
\nonumber \\
&&\left[ \frac{72}{7}\ln \left( 1-\frac{7}{10N_{KK}}\right) -9\ln \left( 1-%
\frac{4}{5N_{KK}}\right) +\frac{9}{7}\ln \left( 1+\frac{1}{2N_{KK}}\right)
\right] ,  \label{4.5}
\end{eqnarray}
where we have used the approximations for the KK masses discussed in the
previous Section. With $N_{KK}=60$ we have $\delta \alpha
_{3}^{-1}(M_{Z})\approx -0.103$, which is indeed a small correction (as it
was expected) and it lies in the range of theoretical uncertainties
estimated in \cite{7}.

However, large uncertainties in the prediction of strong gauge coupling $%
\alpha _{3}(M_{Z})$ might originate from the contributions of the Higgs and
matter KK modes. The point is that, unlike gauge fields, their local kinetic
terms are not controlled by the strong coupling unification assumption. The
corresponding $r$-factors are in general arbitrary parameters and thus the
KK mode spectrum can not be reliably computed. A general treatment of
possible uncertainties is rather difficult since we have not any independent
experimental data to constrain the corresponding KK masses and as well we do
not know any principle how to estimate possible size of the corresponding $r$%
-factors. Instead, just to stress the importance of possible corrections, we
give here some numerical examples assuming certain hierarchies for the $r$%
-factors of different fields.

Consider for example bulk matter fields. Each type of matter fields (with
the same orbifold symmetries) at each KK level $n$ are arranged in the full
SU(5) multiplets, $\overline{5}+10$, see Table 1. Thus if their masses are
degenerate, they will not give any correction at one loop level as it is the
case for the zero modes, i.e. ordinary quarks and leptons in the limit of
vanishing masses. This is indeed the case for the standard KK decomposition
as well. In the general case, it is clear that $(-,+)$ KK matter fields as
well as $(+,-)$ KK matter fields also do not give any extra correction. In
the case of $(-,+)$ the KK mass spectrum is standard since for these fields
both boundaries at $y=0$ and $y=a$ are transparent. Also $(+,-)$ fields feel
only $SU(5)$-invariant fixed point at $y=0$ and thus KK mass spectrum is
SU(5)-invariant. However, different KK modes collected in $(+,+)$ and $(-,-)$
fields might have SU(5) non-universal masses at each KK level due to the
non-universal local kinetic terms at SU(5)-violating fixed point $y=a$. Once
again, $r$-factors for these fields can not be computed within the effective
orbifold field theory and should be treated as independent free parameters.
As a representative example of possible uncertainties in the calculation of $%
\alpha _{3}(M_{Z}),$ consider $\left( -,-\right) $ matter fields. Assume
that $r_{\pi }$-factors for the SU(2)-doublet fields, $10_{Q}^{c\prime }$
and $\overline{5}_{L}^{c\prime }$ are approximately the same and large, $%
\frac{r_{\pi }^{(10_{Q}^{c\prime },\overline{5}_{L}^{c\prime })}}{a}\equiv
\xi _{1}\gg 1$, so that the KK masses can be approximated according to (26),
(27), while SU(2)-singlet fields have small $r_{\pi }$-factors, $\frac{%
r_{\pi }^{(10_{U,E}^{c\prime },\overline{5}_{D}^{c\prime })}}{a}\equiv \xi
_{2}<<1$, so that their KK masses are given by (25). In this case we obtain
the following correction to $\alpha _{3}(M_{Z})$: 
\begin{eqnarray}
\delta \alpha _{3}^{-1}(M_{Z}) &\approx &-\frac{15\eta }{14\pi }\times 
\nonumber \\
&&\left[ \ln \left( \pi N_{KK}\sqrt{\frac{\xi _{1}}{2}}\right)
-\sum_{n=0}^{N_{KK}-1}\ln \left( \frac{n+1}{n+\frac{1}{2}}\right) +N_{KK}\ln
\left( 1+\frac{\xi _{2}}{2}\right) \right] ,  \label{4.6}
\end{eqnarray}
where $\eta $ is the number of matter generations propagating in the bulk.
Taking now, for instance, $\xi _{1}=\frac{1}{\xi _{2}}=50$ and $N_{KK}=60$
as before, we obtain a large correction, $\delta \alpha
_{3}^{-1}(M_{Z})\approx -1.645\eta $, which pushes up the strong gauge
coupling constant to an unacceptable value: 
\begin{eqnarray}
\alpha _{3}(M_{Z}) &\approx &0.148,\text{ for }\eta =1,  \nonumber \\
\alpha _{3}(M_{Z}) &\approx &0.196,\text{ for }\eta =2.  \label{4.7}
\end{eqnarray}
The corresponding correction to the compactification scale $M_{c}$ of (\ref
{4.4b}) is: $\delta \ln \left( \frac{M_{c}}{M_{Z}}\right) \approx 1,38\eta $%
, i.e. 
\begin{eqnarray}
M_{c} &\approx &6\cdot 10^{15}\text{ GeV for }\eta =1,  \nonumber \\
M_{c} &\approx &2.4\cdot 10^{16}\text{ GeV for }\eta =2.  \label{4.8}
\end{eqnarray}
If only the $10$-plets are residing in the bulk, as it is favoured by some
phenomenological considerations \cite{7}, then the correction (\ref{4.6}) is
reduced by anly a factor 0.7 which is still too large. For example, for $%
\eta =2$ and for the above values of $x_{1}$ and $x_{2}$ one obtains $\alpha
_{3}(M_{Z})\approx 0.162$. Alternatively, assuming that $\xi _{1}<<1$ and $%
\xi _{2}>>1,$ the rhs of Eq. (\ref{4.6}) changes the sign (with $\xi
_{1}\leftrightarrow \xi _{2}$) and thus $\alpha _{3}(M_{Z})$ gets
dramatically lowered. We have found large enough corrections for moderate
values of $r$-factors as well.

Certainly, many different numerical examples can be presented which stress
the importance of the correct treatment of KK mass spectrum in the presence
of local kinetic terms. We are not aiming here to give a more general
analysis of possible uncertainties caused by the $r$-factors of different
fields. However, already the above representative numerical examples are
clearly warning us that uncertainties in the calculation of the strong gauge
coupling are actually large and essentially uncontrollable. Strictly
speaking, $r$-factors (especially for the Higgs and matter fields) can
neither be computed or estimated nor can be constrained by the available
experimental data. Thus we are lead to the conclusion that one cannot obtain
reliable predictions in the framework of orbifold GUTs. Nevertheless,
orbifold GUT models could still be viewed as interesting theoretical schemes
where many attractive features, such as GUT symmetry breaking,
doublet-triplet splitting, suppression of the proton decay etc., could have
their counterparts in a more fundamental theory, where, hopefully, the
ambiguity related with local kinetic terms can be also resolved. In this
respect, it seems that many other GUT models which are excluded in the
standard KK decomposition approach might still accommodate the experimental
data within large uncertainties expected in general. A particularly
interesting question is whether one could have low (intermediate) energy
unification in simple orbifold GUTs assuming some particular set of $r$%
-factors, which, in turn, can be verified in high energy experiments.

\section{Summary and outlook}

In this paper we have discussed KK decomposition in higher-dimensional
theories with orbifold compactification. We have shown that the standard KK
decomposition is not valid due to the presence of kinetic terms localized at
the orbifold fixed points. We have also found that the KK mass spectrum and
interactions of KK modes are significantly modified and the phenomenology of
various orbifold models must be reconsidered. As an illustrative example, we
have considered gauge coupling unification in recently proposed realistic 5D
orbifold SU(5) GUT. We have shown that large uncertainties in the low-energy
predictions of the model appear once the local kinetic terms for Higgs and
matter fields are included into consideration and that the predictivity of
the model is essentially lost.

Clearly, our analysis of KK decomposition is relevant as well for a wide
class of other models extensively discussed in the literature. It will be
interesting to see how the supersymmetry and electroweak symmetry breaking
as well as collider phenomenology of particular models with large
compactification radius \cite{28} is affected when the correct KK mode
decomposition we have presented in this work is adopted.

Although we have discussed here the case of a single flat extra dimension,
local kinetic terms are also generally expected when the extra space-time is
not flat, for example, in the case of warped geometry. Thus recently
discussed unification in AdS$_{5}$ \cite{29}, as well as other
phenomenological aspects of the scenario in \cite{30}, might be
significantly modified as well. Finally, it will be certainly interesting to
extend this work to the case of general higher dimensions, where some
non-trivial features might emerge.

\subparagraph{Acknowledgments.}

We thank Z. Berezhiani, H.P. Nilles and Z. Tavartkiladze for discussions.
A.K. would like also to gratefully acknowledge stimulating research
atmosphere at Gran Sasso Summer Institute ''New Dimensions in Astroparticle
Physics'' where this work was finalized.

This work was done by partial financial support of the Academy of Finland
under the Project No. 54023.\newpage

\end{document}